\documentclass[twocolumn]{revtex4b4}
\usepackage{graphicx}

\begin{document}

\bibliographystyle{plain}
%\begin{frontmatter}

\title{The Smectic Phase  of Spherical-Fan Shaped Molecules.
A Computer Simulation Study}

\author{A. G. Vanakaras}
\address{ Department of Materials Science,
          University of Patras,
          Patras 26500, Greece}
\author{D. J. Photinos}
\address{ Depertment of Physics, University of Patras,
          Patras 26500, Greece}
%\corauth[cor]{ Corresponding author,
%               email:photinos@upatras.gr, tel/fax:+3061997461}

\begin{abstract}
We have used Monte Carlo $NPT$ computer simulations to study a
system of spherical fan shaped particles made of three hard discs
fused along a common diameter. The calculated equation of state
indicates a strong, entropy driven, first order transition from
the isotropic phase to a layered phase that has the basic
characteristics of the usual smectic A mesophase but with strongly
correlated rotations about the symmetry axes of neighbouring
molecules.

\end{abstract}
\maketitle

%\begin{keyword}
%     MC simulation, smectic, liquid crystals, non-convex molecules,
%     propellanes, cage molecules
%\end{keyword}

%\end{frontmatter}

During the last decades computer simulations, have become a very
powerful tool not only for testing theories but also for
confirming and predicting phase behaviour in a wide class of
simple as well as complex model systems. In the late 50's, the
fogy landscape in the field of freezing transition for hard sphere
systems was cleared up by the famous computer simulations of Alder
and Wainwright \cite{Alder:JCP57} and of Wood and Jacobson
\cite{Wood:JCP57}. These simulations showed that such systems
could indeed form stable crystals. Three decades later,
simulations by Frenkel and co-workers \cite{Stroobants:PRA87}
confirmed Onsager's theory \cite{Onsager:49} for the entropically
driven order disorder transition in systems of long hard rods.
Since then, numerous computer simulation efforts have shown that
even highly idealized molecular systems, interacting exclusively
with steric repulsions, can exhibit a rich variety of anisotropic
mesophases. Thus, sufficiently anisometric \lq\lq hard" particles
can form phases of the same symmetries with the known liquid
crystalline phases i.e. calamitic and discotic nematics, smectics,
columnar (for a review see \cite{Allen:ACP93}) as well as novel
phases that are yet to be observed in nature (e.g. cubatic
\cite{Veerman:PRA92}). Moreover, computer simulations of hard
particles have provided clear insights into the role of packing
entropy in driving phase transitions to and among liquid
crystalline phases and in determining their structure
\cite{Stroobants:PRA87,Allen:ACP93,Groth:JCP96}. The neglect of
short range \lq\lq soft" interactions or longer ranged
electrostatic interactions or molecular flexibility and the
associated conformational changes, limits  theory and computer
simulations to a qualitative description of real systems, usually
without claims of quantitative reliability. Nevertheless,
mimicking the basic molecular properties with simplified hard body
models and making computer experiments with them, thus taking into
account nearly exactly the short range intermolecular
correlations, could be a first step towards the molecular
engineering of new materials with novel and possibly desirable
properties.

At the same time, the synthetic efforts of chemists in the field
of liquid crystals have produced a number of novel molecular
structures and mesophases that have generated further research
excitement and have inspired new molecular simulations. Most
recently, supermolecular systems of dendritic architecture of
controlable sizes, shapes and rigidity have been made and found to
form mesophases with unusual properties \cite
{Lorenz:AM96,Balagur:JACS97,Richard:LC99,Barbera:CEJ99,Mehl:CC99}.
Interestingly, such systems produce anisotropic fluid phases even
when the global shape of the supermolecules (the \lq\lq molecular
envelope") is not sufficiently (or not at all) anisometric by the
standards used for usual low molar mass liquid crystals
\cite{Barbera:CEJ99,Terzis:MCLC99}. This phenomenon can be
rationalised in terms of strongly anisotropic interactions that
appear at short intermolecular distances where substantial
interpenetration of the molecular envelopes occurs. In hard-body
modelling of the interactions this corresponds to molecules of
highly non-convex shapes, with the non-convexity giving rise to
strong orientational correlations among the molecules at short
distances. Strictly, the shapes of molecules are in general
non-convex. However, orientational ordering in common liquid
crystals is primarily a result of anisometry of the molecular
shape, although non-convexity is clearly involved as an additional
factor in bent-rod molecular models of smectics such as the
zig-zag model \cite{Bartolino:AnP78,Vanakaras:PRE98} and the
banana smectogen models \cite{Camp:JCP99}. Other multi-rod models,
such as the \lq\lq Onsager crosses" studied by Blaak and Mulder
in connection with the formation of cubatic phases
\cite{Blaak:PHD,Blaak:PRE98}, stipulate non-convex molecular
shapes although their primary intention is to convey the
competition among different molecular directions of alignment. To
our knowledge, shape non-convexity has not been explicitly
addressed as the molecular feature underlying the formation of
orientationally ordered fluid phases in classes of compounds where
anisometry of the global molecular shape is marginal or completly
absent.

In this paper we present the first results of our study of liquid
crystalline phases originating from non-convexity of the molecular
shape. To completely remove any influence from global shape
anisometry we have chosen at this stage to study molecules whose
global shape is spherical. We have performed Monte Carlo (MC)
simulations of a system of fan-shaped hard-body molecules made of
three infinitely thin discs fussed symmetrically along a common
diameter, as shown in Fig. 1. The molecular $z$-axis is chosen to
coincide with the direction of the common diameter, the $x$-axis
is chosen along the normal to the surface of one of the three
discs and the $y$-axis completes a right-handed orthogonal
molecular axis frame. The diameter of the discs, and therefore of
the spherical envelope of the entire fan-shaped molecule, is
denoted by $D$.

\begin{figure}
  \begin{center}
    \includegraphics*[width=6cm]{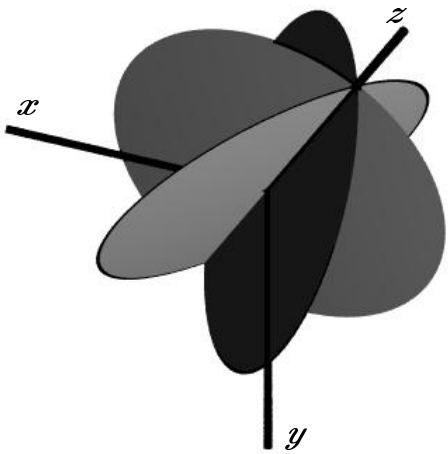}
  \end{center}
 \caption{ Idealised shape of a spherical-fan molecule made of three
           infinitely thin discs fused along a common diameter.}
\end{figure}

Using the standard MC method of Wood \cite{Wood:JCP68} we have
studied the thermodynamic and structural properties of hard
spherical fans in the isothermal-isobaric $(NPT)$ ensemble. An
advantage of this MC-$NPT$ method is that the equation of state of
the system is obtained directly for a sequence of constant
pressures and the corresponding densities are calculated as
ensemble averages. Each MC trial move consists of a random
displacement of the centre of mass of a particle followed by a
random rotation around one of the molecular axes \cite{Allen:B87}.
The move is accepted if the particle in the new state does not
overlap with any of the other particles, otherwise the move is
rejected. The variations in the volume of the sample are
introduced by random changes in the logarithm of the volume
\cite{Frenkel:B96}. The maximum molecular displacements, rotations
and volume changes are adjusted so that the acceptance ratio for
the corresponding trial moves lie around $30\%$. In what follows,
pressure and density are expressed in terms of molecular
quantities according to the usual dimensionless parameterisation:
$P^*=Pv_s/kT$ and $\eta=\rho v_s$, where $\rho=N/V$ is the
molecular number density and $v_s=\pi D^3/6$ is the volume of the
spherical envelope, i.e. of a sphere with diameter equal to the
diameter of the discs. Distances are expressed in units of D.
\begin{figure}
\begin{center}
  \includegraphics[height=6cm,width=8.5cm]{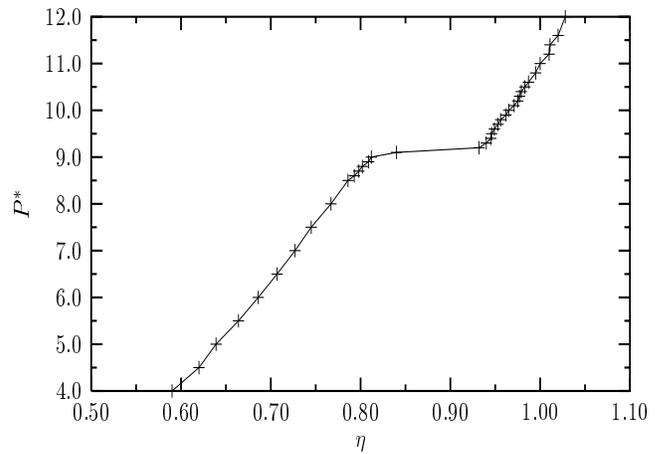}
\end{center}
 \caption{ Calculated $P^*\!-\! \eta$ phase diagram for a system of
 $N=600$ spherical-fan particles as obtained by compression from a
 low pressure isotropic state. Data points obtained by expansion from
 the final, high pressure, state reproduce nearly exactly the same
 phase diagram without notable hysterisis. The solid line is to guide
 the eye.}
\end{figure}

The size of the system and the shape of the simulation box do not
seem to affect critically the results. Simulating systems from
$N=125$ particles up to $N=600$, using cubic or variable shape
box, yield essentially the same results for the equation of state
and the relevant thermal averages, apart from differences in the
standard deviations of the calculated averages. In what follows we
present results of systems containing $N=600$ particles in cubic
boxes. One MC cycle consists on average of N
displacement/reorientation attempts followed by one attempted
volume change. The particles are initially placed on a fcc lattice
and the system is equilibrated at low pressure, well in the
isotropic phase. The system is then gradually compressed to higher
densities by increasing the pressure. For each pressure (using as
initial configuration the last configuration of the previous
pressure) the system is equilibrated for $\sim 5\times10^5$ MC
cycles and ensemble averages are accumulated over at least a
further $\sim 5\times10^5$ MC cycles. The required MC cycles in
the vicinity of the phase transition are by one order of magnitude
longer in order to obtain reliable statistics. The calculated
phase diagram is given in Fig. 2. It is apparent that at $P^*
\approx 9.1$ a first order transition takes place and the system
condenses from a low density phase $\eta \approx 0.83$ to a new
phase with appreciably higher density $\eta \approx 0.93$. Once
the high pressure phase has been reached, the phase diagram is
reproduced with negligible hysteresis by expansion of the system.

\begin{figure}
\begin{center}
  (a)
  \includegraphics*[width=6cm]{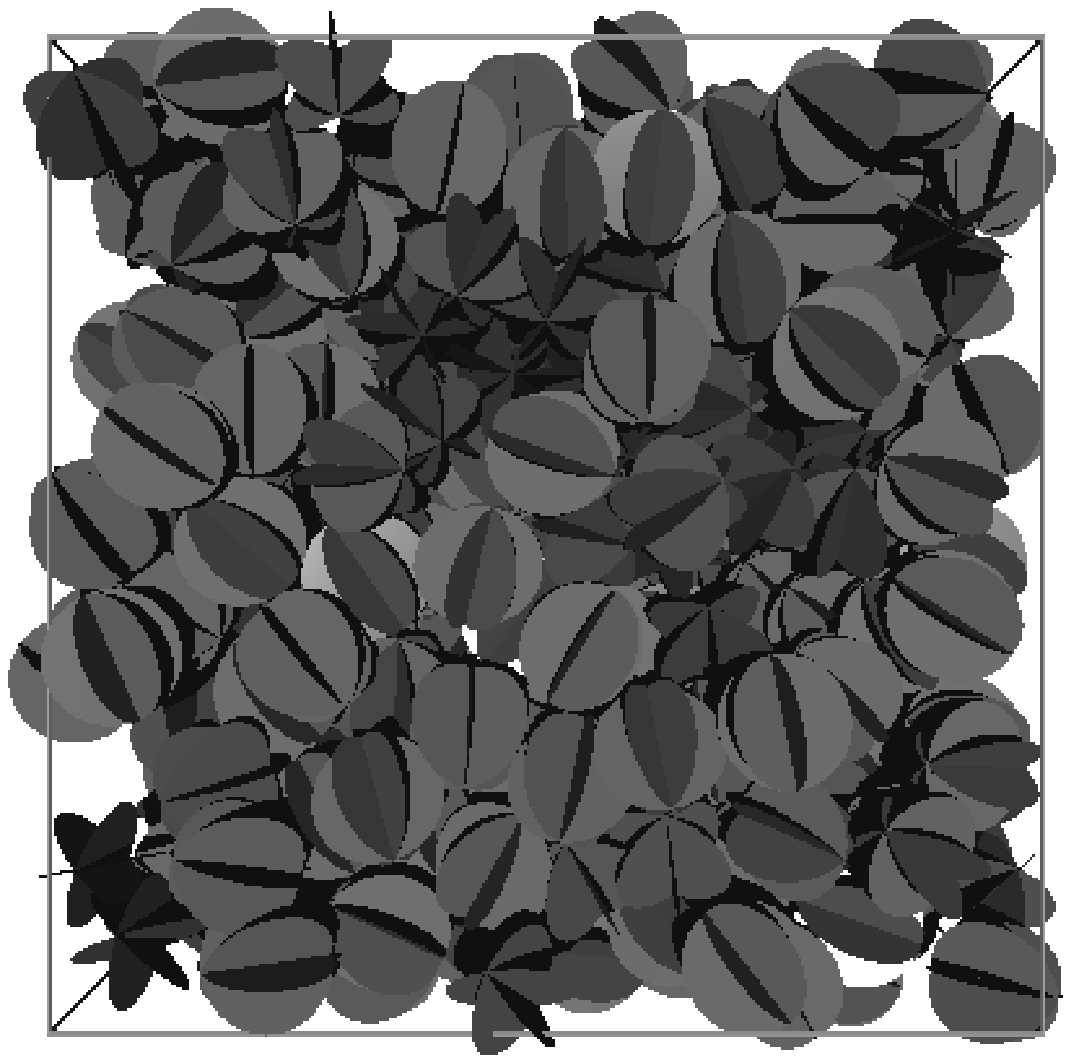} \\
  (b)
  \includegraphics*[width=6cm]{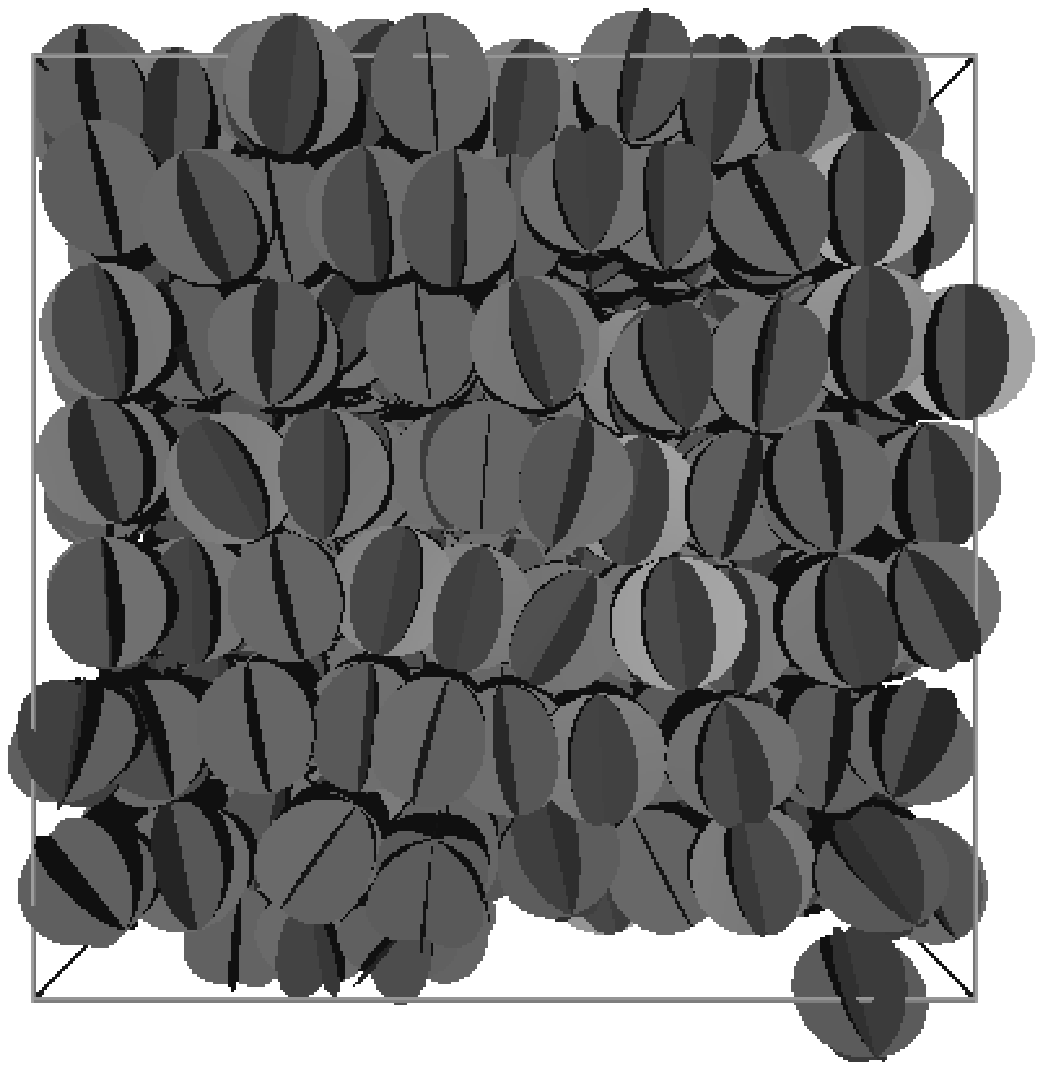}
\end{center}
 \caption{Snapshots of a system of $N=600$ particles;
(a) isotropic phase at $P^*=9.1$ and
(b) layered mesophase at $P^*=9.3$ }
\end{figure}

Visualization of the low and high density phases at the boundaries
of the phase transition, shown in Fig. 3, makes it evident that
the low pressure phase is , as expected, isotropic whereas the
dense phase has both, layered structure and long range
orientational order of the $z$-axes of the spherical fans. In
order to quantify the orientational order we have calculated the
principal order parameter $S=\langle P_2({\bf{z\cdot n}}) \rangle$
where $P_2$ is the second Legendre polynomial, ${\bf{z}}$
corresponds to the direction of the molecular $z$-axis and
${\bf{n}}$ is the director (symmetry axis) of the phase (for
details see \cite{Allen:ACP93,Allen:B87,Zannoni:B79}) and the
angular brackets denote ensemble averaging. In the ordered phase S
increases slowly with increasing the pressure from a value
$S\approx 0.82$ at the transition to $S\approx 0.88$ at $P^*=12$.

In order to characterize the phases in some detail we have
calculated from the simulations the usual orientationally averaged
radial correlation function $g(r)$, as well the orientational
correlation function $g_2^{zz}(r)\equiv\langle
P_2({\bf{z}}(0)\cdot{\bf{z}}(r))\rangle$ where ${\bf{z}}(0)$ is
the direction of the molecular $z$-axis of a particle at the
origin and ${\bf z}(r)$ is the corresponding direction of a
particle at distance $r$ from the origin. Fig. 4(a) shows the
calculated $g(r)$ for four different pressures. For pressures
corresponding to the isotropic phase the first peak appears at
$r=1$ while for higher pressures it appears at $r\approx 0.62$.
The latter value is very near the closest distance of approach
($r=1/\sqrt{3}$) of two molecules oriented such that their
$z$-axes are side-by-side parallel and a disc blade of the one
exactly bisects the dihedral angle formed by two adjacent blades
of the other. The position of this peak, together with the average
intermolecular distance calculated from the density of the ordered
phase and together with analogous features of the other
correlation functions to be discussed below, indicate that the
spherical envelopes of neighbouring molecules overlap considerably
in a way that the blades of one molecule are accommodated in the
grooves between the blades of its side-by-side neighbours, thereby
giving rise to strong correlations of the rotations about the z
molecular axes.  At high pressures, the persisting structure of
$g(r)$ at large distances is primary due to the layering of the
molecules.

In a system with long range orientational order $g_2^{zz}(r)$
becomes long ranged and its limiting value at large distances
tends to $S^2$ \cite{Zannoni:B79}. The first high peak at
$r\approx0.5$ corresponds to the minimum non-overlapping distance
of two particles with parallel $z$-axes (half diameter). Higher
rank orientational correlation functions show the expected
behaviour namely, vanish at long distances in the isotropic phase
and level off at a finite value in the layered phase.

\begin{figure}
\begin{center}
  \includegraphics[width=8cm]{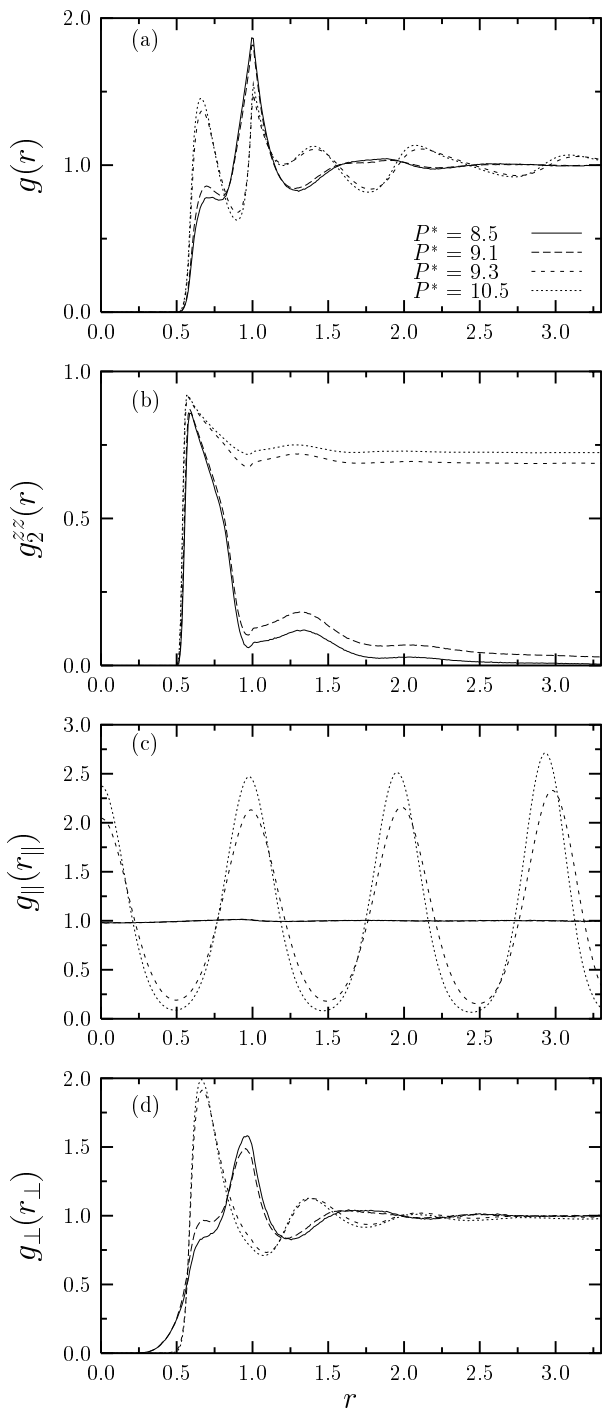}
\end{center}
  \caption{
  Calculated dependence of pair correlation functions on the
intermolecular distance r (in units of the global molecular
diameter D) at various pressures $P^*$. (a) Radial correlation
function $g(r)$, (b) orientational pair radial correlation
function $g_2^{zz}(r)$, (c) longitudinal (along the layer normal)
pair correlation function $g_\parallel(r_\parallel)$ and (d)
parallel (intra-layer)pair correlation function
$g_\perp(r_\perp)$.}
\end{figure}

To examine the positional order we have evaluated the correlation
functions $g_\parallel(r_\parallel)$ and $g_\perp(r_\perp)$ shown
in Fig 4(c) and (d). These describe the probability of finding a
pair of molecules with magnitude $r_\parallel$ ( $r_\perp$) for
the projection of their intermolecular vector along (perpendicular
to) the layer normal. For the layered phase
$g_\parallel(r_\parallel)$ maps the density modulation along the
director, with the period of the density wave yielding the layer
spacing. The nonzero values of the density wave in the inter-layer
region (Fig. 4(c)) indicate that some particles are found between
the layers. The layer spacing at the transition is approximately
one molecular diameter and decreases slightly with increasing
pressure, indicating that the phase does not exhibit strong
interdigitation of the layers , at least up to $P^*=12$.
Furthermore, as evident from Fig. 4(d), the absence of long range
positional correlations (structureless $g_\perp(r_\perp)$ at long
distances) shows that the phase exhibits liquid like positional
behaviour within the layers, characteristic of a smectic A
mesophase. Moreover, we observe that the second high peak in the
radial distribution function, at $r\approx1$, is not exhibited by
$g_\perp(r_\perp)$, indicating that this peak originates from
correlations among particles in different layers.

Spherical fan molecules made of two discs or of more than three
discs have also been considered. Preliminary simulations of two
disc fans showed that their interactions are not sufficiently
anisotropic to clearly produce orientationally ordered phases.
Fans with four or five discs show qualitatively the same behaviour
as three disc fans but with the transition to the ordered phase
shifted towards higher pressures. As the number of discs in the
fan molecules increase, their phase organisation tends to that of
hard spheres. Preliminary runs show that ten disc spherical fans
behave practically as hard spheres. Symmetric three-disc fans with
discs of finite thickness were also considered. The thickness of
the discs controls the extent of side-by-side interdigitation of
the fans and is found to influences critically the phase diagram
and the correlation functions in the ordered phase. Finally,
variants of the three disc spherical fans have also been
considered in which the discs are not fused along a common
diameter; the results will be reported in a forthcoming
communication.

The primitive structures considered in this study are relevant to
certain known classes of organic and inorganic compounds such as
propellanes \cite{Ginsburg:B75} and compounds with cage type
molecular architecture \cite{Tobe:B92}.  Aside from very few known
exceptions \cite{Bassoul:JPC96}, such compounds do not show liquid
crystalline behaviour. With the help of the present simulations
this can understood by simply comparing the effective \lq\lq blade
thickness" to the \lq\lq groove width" of the spheroid fan
molecules and noting that for the molecular interaction to have a
substantial anisotropic component the groves should be
sufficiently deep and wide enough to readily accommodate a blade.
Common carbocyclic fan-shaped compounds do not meet this
requirement since the diameter of the blade-forming rings is not
sufficiently larger than their thickness. The simulated mesophase
behaviour of the thin-disc spherical fan model would therefore be
more directly relevant to supermolecular structures in which the
blade diameter can be made much larger than its thickness while
maintaining sufficient rigidity to confer a spherical fan or cage
geometry to the interacting supermolecules.

{\large{\bf Acknowledgements}} \linebreak
We thank G.H. Mehl and
E.T. Samulski for very helpful discussions. This work was
supported in part by the Greek General Secretariat of Research and
Technology and the European Social Fund under the PENED'99 project
99ED52.

\bibliography{Fan_Refs}

\end{document}